\lineskip=0pt
\parskip=0pt plus 1pt
\hfuzz=1pt
\vfuzz=2pt
\pretolerance=2500
\tolerance=5000
\vbadness=5000
\hbadness=5000
\widowpenalty=500
\clubpenalty=200
\brokenpenalty=500
\predisplaypenalty=200
\voffset=-1pc
\nopagenumbers
\font\bigfont=cmbx12
\font\smallfont=cmr8
\font\smallbf=cmbx8

\def\title#1{\vglue4pc{\exhyphenpenalty=10000\hyphenpenalty=10000
    \baselineskip=18pt
    \raggedright\parindent=0pt
    \bigfont#1\par\rm}\bigskip}
\def\author#1{\vskip1pc
    {\hang\rm#1\par}
     \smallskip}
\def\address#1{{\exhyphenpenalty=10000\hyphenpenalty=10000
    \hang\raggedright\smallfont#1\par\rm}}
\def\abstract{\bigskip\hang{\smallbf Abstract. }\smallfont}
\def\ack{\bigskip\noindent{\bf Acknowledgments\par}%
    \nobreak\noindent\ignorespaces}
\def\refHEAD{\vfill\goodbreak\noindent{\bf References}\par
     \let\ref=\rf
     \nobreak\rm}
\def\references{\refHEAD\parindent=0pt
     \everypar{\hangindent=18pt\hangafter=1
     \frenchspacing\rm}}
\def\rf#1{\par\noindent\hbox to 21pt{\hss [#1]\quad}\ignorespaces}
\def\numrefjl#1#2#3#4#5{\par\rf{#1}#2 {\it #3 \bf #4} #5\par}
\def\numrefbk#1#2#3#4{\par\rf{#1}#2 {\it #3} #4\par}
\def\etal{{\it et al.\ }}
\def\eq(#1){\eqno{(#1)}}
\def\CQG{{\it Class. Quantum Grav.}}
\def\GRG{{\it Gen. Rel. Grav.}}

\def\NC{{\it Nuovo Cimento\/}}

\def\PL{{\it Phys. Lett.}}
\def\PR{{\it Phys. Rev.}}

\title{Can the local energy--momentum conservation laws be derived
solely from field equations?}

\author{Guido Magnano\dag\ and Leszek M Soko\l owski\ddag}

\address{\dag Dipartimento di Matematica, Universit\`a di Torino,
via Carlo Alberto 10, I--10123 Torino, Italy}

\address{\ddag Astronomical Observatory, Jagellonian University,
Orla 171, 30--244 Krakow, Poland}

\abstract
The vanishing of the divergence of the matter stress--energy tensor for
General Relativity is a particular case of a general identity, which
follows from the covariance of the matter Lagrangian in much the same way as
(generalized) Bianchi identities follow from the covariance of the purely
gravitational
Lagrangian. This identity, holding for any covariant theory of gravitating
matter,
relates the divergence of the stress tensor with a combination of the
field equations and their derivatives. One could thus wonder if,
according to a recent suggestion [1], the energy--momentum tensor for
gravitating fields can be computed through a suitable rearrangement of the
matter field equations, without relying on the variational definition. We
show that this can
be done only in particular cases, while in general it leads to ambiguities
and possibly
to wrong results. Moreover, in nontrivial cases the computations turn out
to be more
difficult than the standard variational technique.\rm\bigskip

\def\vol{\sqrt{-g}}
\def\lie#1{{\cal L}_{#1}}

In a recent paper [1] Accioly \etal have observed that for
some well known cases of classical fields interacting with
gravitation (e.g. scalar field, electromagnetic field), upon contracting
the dynamical equation for the matter field with a suitable linear
combination of covariant derivatives of this field, the resulting expression
represents the vanishing of the covariant 4--divergence of some
rank--two tensor. In the cases considered in [1], the latter turns out to
coincide exactly with the stress--energy tensor of the matter field,
according to the usual variational definition.

Let us explain the geometrical origin of this phenomenon. For the
reader's convenience, we recall the derivation of the strong conservation law
(sometimes called ``Bianchi identity for matter'') in the case of metric
theories of gravity, in the form suitable for our aim.  We denote arbitrary
matter
fields by $\psi_A$; the signature of the  metric $g_{\mu\nu}$ is chosen to
be $(-+++)$;
the expressions
$\nabla_{\mu}f$ and $f_{;\mu}$ both denote covariant derivation
relative to the metric $g_{\mu\nu}$ and we set $c=8\pi G=1$. In
accordance with [1], we assume that pure gravitation is described by the
usual Einstein-Hilbert  Lagrangian, although the considerations below
hold, {\it mutatis mutandis,} under more general assumptions. The matter
Lagrangian is
$ L(g,\psi)=L(g_{\mu\nu},R^{\lambda}_{\mu\nu\sigma},\psi_A,\psi_{A;\mu})$:
the dependence on $g$ includes a possible non-minimal coupling to the
curvature. The resulting action is
$$
S=\int_{\Omega}\left({1\over 2}R + L\right)\vol\ d^4x\ . \eq(1)
$$
The (Hilbert) stress-energy tensor and the l.h.s. of the equation
of motion
for the field $\psi_A$ are defined, respectively, as
$$
T_{\mu\nu}(\psi,g)\equiv {-2\over\vol}{\delta (L\vol)\over\delta
g^{\mu\nu}}
\qquad{\rm and}\qquad E^A(\psi,g)\equiv{\partial L\over\partial\psi_A}-
\nabla_{\mu}\left({\partial L\over\partial\psi_{A;\mu}}\right). \eq(2)
$$
The action (1)
is coordinate invariant, so it is left unchanged by any infinitesimal
point transformation $x^{\mu}\mapsto x^{\mu}+\xi^{\mu}(x)$,
with $\xi^{\mu}$
being an infinitesimal vector field [2--4]. For simplicity,
we assume
that $\xi^{\mu}=0$ on the boundary of the integration domain $\Omega$,
so that
the latter is mapped onto itself by the infinitesimal transformation. The
variation of the metric is
$$
\delta g^{\mu\nu}=-\lie{\xi} g^{\mu\nu}=\xi^{\mu;\nu}+\xi^{\nu;\mu},\eq(3)
$$
where $\lie{\xi}$ denotes the Lie derivative. Similarly, for the matter
variables the variation $\delta\psi_A=-\lie{\xi}\psi_A$ can be
expressed [4]
in terms of the covariant derivatives of $\psi_A$, which have the general
form
$$
\nabla_{\mu}\psi_A=\partial_{\mu}\psi_A+{{Z_A}^{\beta}}_{\alpha}(\psi)
\Gamma^{\alpha}_{\mu\beta} \eq(4)
$$
where the coefficients ${{Z_A}^{\beta}}_{\alpha}$ are linear functions of
$\psi$ and depend on its tensorial rank\footnote{$^1$}{For
instance, if $\phi_A$ is a
collection of scalar fields then ${{Z_A}^{\beta}}_{\alpha}\equiv 0$,
if one
deals with a vector field $\psi^{\mu}$ one has ${Z^{\mu\beta}}_{\alpha}
\equiv\psi^{\alpha}\delta^{\mu}_{\beta}$, and so on.}; namely, one has
$$
\lie{\xi}\psi_A\equiv\xi^{\alpha}\nabla_{\alpha}\psi_A-
{{Z_A}^{\beta}}_{\alpha}\nabla_{\beta}\xi^{\alpha}. \eq(5)
$$
The invariance of the action implies
$$
0 = \delta S =\int_{\Omega}\vol\left[{1\over 2}(G_{\mu\nu}-T_{\mu\nu})
\delta g^{\mu\nu}+ E^A\delta\psi_A\right]d^4x \eq(6)
$$
plus a surface integral which vanishes under our assumptions on $\xi^{\mu}$.
Using (3) and (5), dropping again a total divergence and taking
into account the
(metric) Bianchi identity $\nabla^{\nu}G_{\mu\nu}\equiv 0$,
the integrand in (6) becomes
$
\xi^{\mu}[\nabla^{\nu}T_{\mu\nu}-E^A\nabla_{\mu}\psi_A-
\nabla_{\beta}(E^A{{Z_A}^{\beta}}_{\mu})].
$
The vector field $\xi^{\mu}$ being arbitrary in the interior of $\Omega$,
the vanishing of the integral (6) entails the following identity:
$$
E^A\nabla_{\mu}\psi_A=\nabla^{\nu}(T_{\mu\nu}-E^AZ_{A\nu\mu}),\eq(7)
$$
which implies the local conservation law $\nabla^{\nu}T_{\mu\nu}=0$
for solutions of the field equation $E^A(\psi,g)=0$.

Equation (7), which is quoted in this form in [4]
but was essentially contained in earlier work (see e.g.~Ref.~[2]), shows that
contracting the free index of the l.h.s.~$E^A$ of the  matter field
equation with $\nabla_{\mu}\psi_A$ always yields the full
divergence of a rank-two tensor. This is a universal property, holding
for {\it any} matter and {\it any} (generally covariant)
Lagrangian.

Apparently, (7) provides a possible way to compute the tensor
$T_{\mu\nu}$ directly from the field equations, and therefore to obtain the
r.h.s.~of the Einstein equation without having to deal with the action
principle. This possibility has been advocated by the authors of [1], who
however
identify directly the stress--energy tensor with the {\it full\/} expression
occurring in the divergence on the r.h.s. of (7), thus overlooking the term
$\nabla^{\nu}(E^AZ_{A\nu\mu})$. Of course, this term vanishes for
solutions of the field equation $E^A=0$, but for most purposes the correct
stress--energy tensor of the theory should be unambiguously defined for {\it
any\/} field configuration, and not only for exact solutions. Only for the
examples considered in [1] can the correct stress--energy tensor be
obtained according to their prescription -- in the first two examples (scalar
massive Klein--Gordon field and non--minimally coupled scalar field), just
because for a scalar field this term vanishes.  For the subsequent example,
i.e.~for the electromagnetic potential $A_{\mu}$ coupled (either minimally or
non-minimally) to gravity, the identity (7) becomes
$$
E^{\nu}A_{\nu;\mu}=\nabla^{\nu}(T_{\mu\nu}+E_{\nu}A_{\mu}). \eq(8)
$$
If one multiplies Maxwell
equations ${F^{\mu\nu}}_{;\nu}=0$ by $A_{\mu;\alpha}$, after a number of
rearrangements one arrives at
$$
{\tau^{\mu\nu}}_{;\nu}=R^{\mu\nu\alpha\sigma}F_{\nu\alpha}A_{\sigma},
\qquad{\rm where}\quad
\tau^{\mu\nu}\equiv -{A_{\alpha}}^{;\mu}F^{\nu\alpha}+
{1\over 4}g^{\mu\nu}F_{\alpha\beta}F^{\alpha\beta}
$$
is the {\it canonical\/} energy-momentum tensor, which is
non-symmetric and
gauge-dependent and as such has no physical meaning. This may be
particularly
confusing if one makes this computation in
flat space, where this tensor is conserved.
However, the authors of [1] multiply instead the electromagnetic
equation $E^{\nu}=0$ by
$F_{\mu\nu}=\nabla_{\mu}A_{\nu}-\nabla_{\nu}A_{\mu}$,
thus producing an additional
term which cancels exactly the last term on the r.h.s. of (8): in fact,
$E^{\nu}F_{\mu\nu}=E^{\nu}A_{\nu;\mu}-\nabla_{\nu}(E^{\nu}A_{\mu})+
A_{\mu}\nabla_{\nu}E^{\nu}$ and the divergence of the field equation,
$\nabla_{\nu}E^{\nu}$, vanishes {\it identically\/} in the case
discussed. Unfortunately, in general it is not possible to get rid of the
additional term in (7) by contracting the field equation with a suitably
modified
combination of covariant derivatives of the field: namely, the prescription
fails
to work whenever
$\nabla_{\nu}E^{\nu}$ does not vanish identically.
For example, consider a vector field $A_{\mu}$ with a Lagrangian
$$
L=S_{\mu\nu}S^{\mu\nu},\qquad{\rm where}\qquad
S_{\mu\nu}=A_{\mu;\nu}+A_{\nu;\mu}=S_{\nu\mu}. \eq(9)
$$
The Euler-Lagrange field equations are
Maxwell-like, ${S^{\mu\nu}}_{;\nu}=0$, but  now
${S^{\mu\nu}}_{;\mu\nu}$ does not
vanish for arbitrary $A_{\mu}$. Thus, the r.h.s.~of (8)
includes a term, $A_{\mu}{S^{\lambda\nu}}_{;\lambda\nu}$,
which contains linearly the
third derivatives of the field $A_{\mu}$ and therefore
{\it cannot\/} be cancelled by any
linear combination of the form
$c_{\lambda\mu}^{\alpha\beta}A_{\alpha;\beta}E^{\lambda}$
(with $c_{\lambda\mu}^{\alpha\beta}$ constant coefficients)
which would contain only the first and second derivatives.

We further remark that any manipulation of the equations along the ideas
presented in [1], aimed at producing the correct result in at least a
reasonably wide class of theories, is likely to exploit the freedom of
multiplying the field equations by a constant: then, the resulting
stress-energy tensor would be
determined up to a  multiplicative factor (of any sign). However, to
identify $T_{\mu\nu}$ in the Einstein field equation with the physical
energy and momentum density, the correct numerical factors (which are
suppressed in many papers) should be included in matter Lagrangians.
This is the most elementary reason why the information contained in the
Lagrangian cannot be fully replaced by the knowledge of the matter field
equations. The actual risk inherent in playing with the field equations alone
is well illustrated by a computation given in [1] for the non-minimally
coupled scalar field. The minimal coupling, according to their equation
(8)[1], should be recovered when $f(\phi)$ reduces to any constant.
However, the terms due to the non-minimal coupling in the stress tensor
given on p.~1165 disappear {\it only\/} if
$f(\phi)=0$. In this case, however, the stress-energy tensor reduces to
exactly {\it half\/} of the expression (7)[1] given for the minimally
coupled field. Setting instead $f(\phi)=1/(2\kappa)$, as is suggested in
[1], leads to a {\it wrong\/} result: the Einstein equation becomes the
correct one (multiplied by ${1\over 2}$), but the expression for $T^{\mu\nu}$
equals (7)[1] {\it only\/} for solutions of the Einstein equation.

We shall discuss below whether a more refined prescription might be devised,
taking into account the additional term $E^AZ_{A\nu\mu}$ in (7), to implement
effectively the proposal of [1]; before that, let us address a side
aspect which may also cause misunderstanding.

The authors of [1] make a
preliminary distinction between the theories involving only minimal
coupling and
those including non--minimal coupling. In doing so, they assume that minimal
coupling occurs if the  matter Lagrangian contains only the metric $g$ and not
the Levi--Civita connection $\Gamma$, which is incorrect. The minimal coupling
rule  states that in a freely falling local reference frame (i.e.~in a locally
geodesic coordinate system) the matter Lagrangian should reduce to the form of
the flat-space matter Lagrangian.  This implies that a minimally coupled matter
Lagrangian should  not contain {\it curvature\/} terms; however, it will
necessarily contain  first derivatives of the matter fields, and in general
this
entails the occurrence of  Christoffel symbols to
ensure covariance. In the cases of scalar fields and of electromagnetism it is
possible to get rid of the Christoffel connection in the
matter Lagrangian, but in other cases a ``minimally coupled Lagrangian" not
including $\Gamma$ does not exist at all.
Examples are provided by spin-${1\over 2}$ Dirac field [5],
linear spin-two fields in a Ricci-flat
background [3], spin-${3\over 2}$ field [6,7] and the vector field (9)
considered above.
It is worth mentioning here that there might exist cases where the
metric connection $\Gamma$ occurs in the matter Lagrangian not
only through
covariant derivatives of the fields. Consider for instance a
matter field
$\gamma^{\alpha}_{\mu\nu}$
which transforms as a (non-metric) linear
connection. Then one defines a tensor
$Q^{\alpha}_{\mu\nu}\equiv\gamma^{\alpha}_{\mu\nu}-
\Gamma^{\alpha}_{\mu\nu}$
and expresses the matter Lagrangians in terms of $Q$ and its covariant
derivatives (with respect to $\Gamma$). Should such a coupling be considered
a non--minimal one?

In any case, we stress that whether the matter fields are coupled to gravity
minimally or non--minimally is irrelevant by itself in this context. What
really
affects the discussion is whether the full Lagrangian can be split into a pure
gravitational part plus a matter term, or not; if it cannot, the stress--energy
tensor itself is not well--defined. Now, non--minimal coupling does not
necessarily prevent such a splitting of the Lagrangian. However, for some
non-minimally coupled theories one is led to define an {\it effective\/}
stress-energy tensor (for instance in scalar--tensor teories; see Ref.~[8],
eq.~(2.25)--(2.28)\footnote{$^2$}{In [8], eq.~(2.25)  is affected by
a misprint and should read
$G_{\mu\nu}=\cdots\equiv
\theta_{\mu\nu}(\phi,g)+{8\pi\over\phi}T_{\mu\nu}(\Psi,g)$.}) which does {\it
not\/} correspond to a separate matter Lagrangian; as long as
one looks at the gravitational equations, this tensor is
indistinguishable from a genuine variational stress--energy tensor,  yet in
this
case trying to derive the proper conservation laws from the field equations
would lead to ambiguities.

A different, but not unrelated, problem is connected with the fact that while
dealing with gravitational theories including non-minimal coupling,
possible redefinitions of the fields (including the metric) are often
considered in  the literature
(see for instance [9], where  a particular
application of the general method described in [10] is
discussed in detail);
this causes further ambiguities, on completely different grounds,
on the definition of the
correct physical stress-energy tensor [11--13]. Such
ambiguities can be removed
by a careful analysis of the Lagrangian formulation of the
model [8], while a prescription based on the field equations alone would be
unsuitable for a rigorous  approach to this problem.

Let us finally come back to the main question: does the universal formula (7)
allow us to obtain a practical method to compute the stress--energy tensor?
The identity (7) allows to single out $T_{\mu\nu}$ only up to
the addition of an arbitrary
tensor being identically divergenceless, for instance
$V^{\mu\nu}=\nabla_{\alpha}W^{\mu\nu\alpha}$, where
$W^{\mu\nu\alpha}=W^{[\mu\nu\alpha]}$
is any totally antisymmetric tensor (then Ricci and Bianchi
identities imply
$\nabla_{\nu}V^{\mu\nu}\equiv 0$).
The procedure should then at least be supplemented by the
requirement that the
resulting rank-two tensor be symmetric; yet, in general it is
unclear whether this
would ensure uniqueness. The uniqueness cannot be restored
by requiring that the
kinetic part of $T_{\mu\nu}$ be quadratic in the field variables,
because for a non-minimal coupling in general the variational
stress tensor
contains terms linear in the highest derivatives [8].
Furthermore, the
method remains highly non-algorithmic, not because the
manipulation to
be performed on the field equation would be only vaguely
defined (as it appears in Ref.~[1]) but rather because recasting the
l.h.s~of (7) into
the  divergence
of a symmetric tensor needs several non-trivial tricks to be found
{\it ad hoc.}

For instance, the computation for the vector field occurring in (9)
using (8) would be quite tricky; it is much easier to use a
generic formula for
$A_{\mu}$ following from (2),
$$
T_{\mu\nu}(A,g)=Lg_{\mu\nu}-2{\partial L\over\partial g_{\mu\nu}} +
2\nabla_{\beta}\left(
Q^{(\alpha\beta)}A_{(\mu}g_{\nu)\alpha}\right)-
\nabla_{\lambda}\left(
Q^{(\alpha\beta)}A^{\lambda}g_{\mu(\alpha}g_{\beta)\nu}\right),
\eq(10)
$$
where $Q^{\mu\nu}\equiv \partial L/\partial A_{\mu;\nu}$.

For a spin-two field, which is represented by a
symmetric rank-two tensor $\psi_{\mu\nu}$, the identity (7) reads
$$
{T^{\mu\nu}}_{;\nu}= -2{\psi^{\mu}}_{\alpha}{E^{\alpha\nu}}_{;\nu}
-2E^{\alpha\beta}\left({\psi^{\mu}}_{(\alpha;\beta)}-{1\over 2}
{\psi_{\alpha\beta}}^{;\mu}
\right), \eq(11)
$$
and $E^{\mu\nu}$ is so complicated [3] that
guessing the appropriate manipulations
to get $T_{\mu\nu}$ from (11)
would hardly be successful.

We conclude that to {\it compute\/} the stress-energy tensor it is
in general both safer and easier to rely on the variational
definition (2),
which provides an algorithmic and unique prescription. The
computation can be performed almost straightforwardly using
standard computer packages for tensor calculus (see e.g.~[14]).
On the other hand, if the procedure suggested in [1], rather than a purely
computational trick, is intended to provide an alternative way
to {\it define\/} the matter source term in a general-relativistic
gravitational theory, circumventing the need to introduce a
Lagrangian density
for the model, then the whole approach is misleading. In fact, the
definition itself of stress--energy tensor, as well as the property
that the matter field equation can be recast into a full
divergence by a suitable manipulation, rely on the existence of an action
principle from which both the gravitational (e.g. Einstein's) equation and
the matter
field equation should be derived.

\ack
This work is sponsored by the MURST National Project {\it ``Metodi
Geometrici e Probabilistici in Fisica Matematica''}.
The work of LMS was partially supported by the grant
KBN no.~2-P03B-011 13.
GM thanks the Institute for Theoretical Physics of the
Jagellonian University
for hospitality in Krakow.

\references

\numrefjl{1}{Accioly A J, Azeredo A D, de Arag\~ano C M L
and Mukai H 1997}{\CQG}{14}{1163}

\numrefbk{2}{Trautman A 1962 {\it Conservation Laws in
General Relativity,}
in:}{Gravitation, an introduction to current research}
{ed L Witten (New York, Wiley) p~169}

\numrefjl{3}{Aragone C and Deser S 1980}{\NC}{57B}{33}

\numrefjl{4}{Ferraris M and Francaviglia M 1992}{\CQG}{9}{S79}

\numrefbk{5}{DeWitt B S 1965}{Dynamical Theory of Groups and Fields}
{(New York, Gordon \&\ Breach)}

\numrefjl{6}{Ferrara S, Freedman D and van Nieuwenhuizen P
1976}{\PR}{D 13}{3214}

\numrefjl{7}{Deser S and Zumino B 1976}{\PL}{B 62}{335}

\numrefjl{8}{Magnano G and Soko\l owski L M 1994}{\PR}{D 50}{5039}

\numrefjl{9}{Accioly A J, Wichoski U F, Kwok S F
and Pereira da Silva N L P 1993}{\CQG}{14}{L215}

\numrefjl{10}{Magnano G, Ferraris M and Francaviglia M
1987}{\GRG}{19}{465}

\numrefjl{11}{Brans C H 1988}{\CQG}{5}{L197}

\numrefjl{12}{Soko\l owski L M 1989}{\CQG}{6}{2045}

\numrefjl{13}{Ferraris M, Francaviglia M and Magnano G
1990}{\CQG}{7}{261}

\numrefbk{14}{Parker L and Christensen S M 1994}{MathThensor:
a System for Doing
Tensor Analysis by Computer}
{(Reading, Addison-Wesley)}

\bye